\documentclass[showpacs,preprintnumbers,amsmath,amssymb]{revtex4}
\usepackage{amsmath}
\usepackage{graphicx}
\usepackage{subfig}
\usepackage{hyperref}
\usepackage{amssymb}
\usepackage[english]{babel}
\usepackage{epsfig}
\usepackage{wasysym}
\usepackage{bm}
\usepackage{color}
\usepackage{epsfig}
\usepackage{amsfonts}
\usepackage{graphicx}
\usepackage{dcolumn}
\usepackage{indentfirst}
\usepackage{float}

\usepackage[utf8]{inputenc}

\begin{document}
\title{Generalized models for black-bounce solutions in $f(R)$ Gravity\\}
\author{J\'{u}lio C. Fabris$^{(a,b)}$}\email{fabris@pq.cnpq.br}
\author{Ednaldo L. B. Junior$^{(c)}$}\email{ednaldobarrosjr@gmail.com}
\author{Manuel E. Rodrigues$^{(d,e)}$}\email{esialg@gmail.com}

\affiliation{$^{(a)}$ Universidade Federal do Esp\'{\i}rito Santo, CEP 29075-910, Vit\'{o}ria/ES, Brazil}

\affiliation{$^{(b)}$ National Research Nuclear University MEPhI, Kashirskoe sh. 31, Moscow 115409, Russia}

\affiliation{$^{(c)}$ Faculdade de F\'{\i}sica, 
Universidade Federal do Par\'{a}, Campus Universitário de Tucuru\'{\i}, CEP: 
68464-000, Tucuru\'{\i}, Par\'{a}, Brazil}

\affiliation{$^{(d)}$ Faculdade de Ci\^{e}ncias Exatas e Tecnologia, 
Universidade Federal do Par\'{a}\\
Campus Universit\'{a}rio de Abaetetuba, CEP 68440-000, Abaetetuba, Par\'{a}, 
Brazil}

\affiliation{$^{(e)}$ Faculdade de F\'{\i}sica, PPGF, Universidade Federal do 
 Par\'{a}, 66075-110, Bel\'{e}m, Par\'{a}, Brazil}


%
\begin{abstract}

 In this article, the implementation of black-bounce solutions in $f(R)$ theories is investigated. Black-bounce solutions are regular configurations of the static spherically symmetric space-time, containing both black holes and wormholes structures. In General Relativity (GR), black-bounce solution implies violation of the energy conditions. We investigate the same issue in $f(R)$ theories using two strategies: first, supposing a given form for the $f(R)$ function and then determining the matter behavior; second, imposing a condition on the matter density and obtaining the resulting $f(R)$ function. In all cases, a given structure for the metric functions is supposed. Violation of the energy conditions still occur but they are less severe than in the corresponding GR cases. We propose a zero-density model that has horizons, which differs from the GR case. We also propose a model with positive energy density and show that $\rho+p_r>0$, which was not the case in GR.

\end{abstract}

\maketitle


\section{Introduction}
\label{sec1}
Black-bounces spacetimes \cite{Visser2} combines two important scenarios in the study of extreme astrophysical objects: regular black holes \cite{bar,h} and traversable wormholes \cite{wh1,visser, wh2,wh3}.
Even if the singularity presented in black holes, like the Schwarzschild one, is covered by an event horizon, the possible existence of a singular region, where any
physical description is impossible, remains a disturbing feature pointing to an incompleteness of the underlying theory, in occurrence the General Relativity (GR) theory.
Hence, the search for regular configurations, including in the inner region, for black holes may lead to indications on how GR may be extended or modified in order to have a more consistent theory of gravity. On the other hand, even if wormholes remain a pure hypothetical object, the conditions for their possible realization in nature consist a
very relevant topic of research. In particular a lot of interest is devoted to traversable wormholes. It is quite possible that exotic matter, with violation of at least some energy conditions, is necessary to have a regular black hole and a traversable wormhole \citep{regularBH, LOBO}. Black-bounces try to address both problems (regular black holes and traversable whormhole) as particular cases of a unique general configuration.

In its simplest formulation, the black-bounces approach use the inverse problem: a given solution is supposed, with the desired features, and the conditions of its realization are
identified. For example, the usual Schwarzschild solution,
\begin{eqnarray}
\label{sch}
ds^2 = \biggr(1 - \frac{2m}{r}\biggl)dt^2 - \biggr(1 - \frac{2m}{r}\biggl)^{-1}dr^2 - r^2d\Omega^2.
\end{eqnarray}
and the Ellis-Bronnikov wormhole, 
\begin{eqnarray}
\label{eb}
ds^2 = dt^2 - dr^2 - (r^2 + L^2)d\Omega^2.
\end{eqnarray}
may be, in a unified way, replaced by, 
\begin{eqnarray}
\label{bb1}
ds^2 = \biggr(1 - \frac{2m}{\sqrt{r^2 + L^2}}\biggl)dt^2 - \biggr(1 - \frac{2m}{\sqrt{r^2 + L^2}}\biggl)^{-1}dr^2 - (r^2 + L^2)d\Omega^2.
\end{eqnarray}
The parameter $L$ with dimension of length is introduced ad hoc in order to avoid the singularity, using the transformation $r \rightarrow \sqrt{r^2 + L^2}$. Depending on the relative values of $m$ and $L$, regular or traversable wormholes can be obtained. It is important to remark that, in opposition to some other similar approaches to regularize the black hole singularity, in the black-bounce proposal the radial function is also transformed what allows to obtain wormhole-like configurations.

The modification described above implies that the metric \eqref{bb1} is not anymore a vacuum solution as the Schwarzschild solution. In fact, it is necessary to have some kind of fluid to support the solution \eqref{bb1}. In the spherically symmetric and static phase, the fluid must have a tangential $p_t$ and a radial pressure $p_r$. The conditions to the existence of black-bounce configurations may imply that some or even all energy conditions are violated. This is common feature in the study of regular black holes and traversable wormholes as already pointed out.

In this work, we consider the conditions matter must obey in order the metric \eqref{bb1} to be a solution of the $f(R)$ theory, which is a non-linear generalization of the Einstein-Hilbert Lagrangian. The gravity theories based on the $f(R)$ proposal have received a lot of attention recently in connection with the problem of the dark sector of the universe
\cite{felice}. We remark {\it en passant} that, through a field redefinition, a $f(R)$ theory may be recast as a Brans-Dicke theory with $\omega = 0$ and a given potential \cite{bd}. In the scalar-tensor version the coupling to matter allows to implement in a suitable way a chameleon mechanism. From the cosmological point of view, only some classes of $f(R)$
functions lead to viable models. A viable $f(R)$ theory must satisfy some conditions, for example $f_R$ and $f_{RR}$ (the subscript indicating derivatives with respect to $R$) must be positive, assuring a positive sound speed and the absence of ghosts \cite{felice}. In cosmology it is required also that a matter dominated epoch must exist and to correspond to a saddle point in the phase space, lasting enough in order to assure the formation of structures \cite{felice}. In the analysis to be carried out here, we are interested in the possibility to have the solution \eqref{bb1} in such a way that the violation 
of the energy conditions, which plagues the original black-bounce formulation in GR, can be at least alleviated. 

In the context of $f(R)$ gravity there are a few approaches to the study of black holes \cite{bh1,bh2,bh3,bh4,bh5,bh6,bh7,bh8,bh9,bh10,bh11}. In the direction of regular black holes, there are, to our knowledge, only two works \cite{rbh1,rbh2}. As for wormholes, there are many papers that address various physical characteristics, among them, we can mention \cite{woh1,woh2,woh3,woh4,woh5,woh6,woh7}. 

In order to perform a concrete analysis, a given form to the $f(R)$ function must be specified. Two polynomial forms of the $f(R)$ function will be taken into account. A third model, where the function is exponential, will also be analyzed. These models satisfy, in principle, the requirements for a viable $f(R)$ theory. The polynomial function may be considered as a truncation of the general exponential function. The conditions to have the black-bounce solution, either black hole or wormhole ones, will be taken into account.

Our main results are the following: some models have positive density in some region of space-time; there is a model with zero energy density outside the event horizon; we find also a model that satisfies $\rho + p_r > 0$ outside the event horizon, and $\rho > 0$ in all spacetime, a situation which never happens in the GR context. Also, $SEC_3$, the strong energy condition, is violated for some cases.

The article is organized as follows. In next section, the relevant equations for the $f(R)$ gravity are set out. The new black-bounce solutions in $f(R)$ gravity are presented in
section \ref{sec2} for the three models described above. In section \ref{sec3}, 
we present the quadratic model in \ref{subsec1}, the zero density model in \ref{subsec2}, and finally, in \ref{subsec3}, the positive density model. Our final considerations are presented in \ref{sec4}.


\section{The equations of motion in $f(R)$ Gravity}
\label{sec2}
The General Relativity (GR) theory is based on the Einstein-Hilbert Lagrangian action,
\begin{eqnarray}
S_{EH}=\int d^4x\sqrt{-g}\left[R+2\kappa^2\mathcal{L}_m\right],
\label{action}\;,
\end{eqnarray}
where $g$ stands for the determinant of the metric $g_{\mu\nu}$, $\kappa^2=8\pi G/c^4$, with $G$ and $c$ being the Newton's gravitational constant and the speed of light, respectively
(from now on we choose units such that $G=1$ and $c=1$), $R$ is the Ricci scalar and $\mathcal{L}_m$ represents the 
Lagrangian density of matter and other fields.

One of the possible generalizations of GR is made by replacing the curvature scalar by any nonlinear function $f(R)$ of the Ricci scalar such that, in the metric formalism, the equations of motion are obtained from the action,
\begin{eqnarray}
S_{f(R)}=\int d^4x\sqrt{-g}\left[f(R)+2\kappa^2\mathcal{L}_m\right]
\label{action}\;,
\end{eqnarray}


Applying the variational principle in terms of the metric to the action (\ref{action}), we find the following field equations: 
\begin{eqnarray}
\label{fe}
f_RR^{\mu}_{\;\;\nu}-\frac{1}{2}\delta^{\mu}_{\nu}f+\left(\delta^{\mu}_{\nu} 
\square 
-g^{\mu\beta}\nabla_{\beta}\nabla_{\nu}\right)f_R=\kappa^2\Theta^{\mu}_{
\;\;\nu}\label{eqfR}\;,
\end{eqnarray}  
where $f_R\equiv df(R)/dR$, $R^{\mu}_{\;\;\nu}$  is the Ricci 
tensor, $\nabla_{\nu}$ stands for the covariant 
derivative, $\square\equiv 
g_{\alpha\beta}\nabla_{\alpha}\nabla_{\beta}$ is the d'Alembertian, and 
$\Theta_{\mu\nu}=-\frac{2}{\kappa^2\sqrt{-g}}\frac{\delta(\sqrt{-g}\mathcal{L}_m)}{\delta(g^{\mu\nu})}
$ 
is the matter energy-momentum tensor.  
\par 
In the present work, we will analyze the black-bounce space-time and for that 
we then consider a spherically symmetric and static space-time, whose element 
line reads
\begin{eqnarray}
ds^2=e^{a(r)}dt^2-e^{b(r)}dr^2-\Sigma(r)^2\left[d\theta^2+\sin^2\theta 
d\phi^2\right] 
\label{ele}\;,
\end{eqnarray} 
where $a(r)$, $b(r)$ and $\Sigma(r)$ are arbitrary functions of the radial coordinate 
$r$. The determinant of the metric is $g=-e^{a(r)+b(r)}\Sigma(r)^4\sin^2\theta$. We can model the material content by an anisotropic fluid, so the energy-momentum tensor becomes $\Theta_{\;\;\nu}^{\mu}=diag\lbrace \rho(r), -p_r(r), -p_t(r), -p_t(r)\rbrace$. The equations of motion for the $f(R)$ gravity  are then 
found by introducing the line element \eqref{ele} in the field equations \eqref{fe}:
\begin{eqnarray}
&&\frac{e^{-b}}{4\Sigma}\Big\{4\Sigma\frac{d^2f_R}{dr^2}+ 
2\left[4\Sigma'-\Sigma b'\right]\frac{df_R}{dr}+\big[\Sigma\left(a'b'-2a''-(a')^2\right)-4a'\Sigma'\big]f_R+2\Sigma e^bf\Big\}=-\kappa^2\rho ,\label{eq1}\\
&& \frac{e^{-b}}{4\Sigma}\Big\{2\left[4\Sigma'+\Sigma a'\right]\frac{df_R}{dr}+ 
\left[(4\Sigma'+\Sigma a')b'-2\Sigma a''-\Sigma(a')^2-8\Sigma''\right]f_R+2\Sigma e^bf\Big\}=\kappa^2p_r,\label{eq2}\\
&&\frac{e^{-b}}{2\Sigma^2}\Big\{2\Sigma^2\frac{d^2f_R}{dr^2}+\left[\Sigma^2(a'-b')+2\Sigma\Sigma'\right]
\frac{df_R}{dr}+\left[\Sigma(b'-a')\Sigma'+2\left(e^b-(\Sigma')^2-\Sigma\Sigma''\right)\right]f_R+\Sigma^2e^bf\Big\} =\kappa^2p_t ,\label{eq3}
\end{eqnarray}
where the prime ($'$) stands for the total derivative with respect to the 
radial coordinate $r$. 
\par
The Ricci scalar, for the present case, is given by, 
\begin{eqnarray}
R=\frac{e^{-b}}{2\Sigma^2}\left[\Sigma^2\left((a')^2-a'b'+2a''\right)+4\Sigma\left((a'-b')\Sigma'+2\Sigma''\right)+4(\Sigma')^2-4e^b\right]\,. \label{Rscal}
\end{eqnarray}
The Kretschmann scalar is given by
\begin{eqnarray}
K=R^{\mu\nu\alpha\beta}R_{\mu\nu\alpha\beta}=\frac{e^{-2 b} }{4 \Sigma^4}\Big[8 \Sigma^2 \left((a')^2 (\Sigma ')^2+\left(b' \Sigma'-2 \Sigma ''\right)^2\right)+\Sigma^4
   \left(-a' b'+(a')^2+2 a''\right)^2+16 \left(e^{b}-(\Sigma ')^2\right)^2 \Big]\label{k}\,.
\end{eqnarray}

For a solution to be regular $\Sigma$ must be non-zero, $\Sigma'$ and $\Sigma''$ must be finite everywhere. Moreover, the metric functions $e^{a(r) }$ and $e^{b(r)}$ as well as their first and second derivatives must be finite in all space.

In order to perform an analysis of the physical properties of this class of solution, we must take into account the energy condition relations for the $f(R)$ theory. The energy condition for this theory will be written using the relations for $\rho$, $p_r$ and $p_t$\cite{visser}:

\begin{eqnarray}
&&NEC_{1,2}(r)=SEC_{1,2}(r)=WEC_{1,2}(r)=\rho+p_{r,t}\geq 0\;,\label{cond1}\\
&&SEC_3(r)=\rho+p_{r}+2p_{t}\geq 0\,,\label{sec}\\
&&DEC_{1,2}(r)=\rho- |p_{r,t}|\geq 0, \label{cond3a}\\ &&WEC_3(r)=DEC_{3}(r)=\rho\geq \label{conddec3}
0\;.\label{cond3}
\end{eqnarray}
The indices $1$ and $2$ indicate the presence of the radial or tangential pressure, respectively, while the indice $3$ indicates the presence or absence of all pressure terms.
Remark the identities $NEC_{1,2}\equiv WEC_{1,2}\equiv SEC_{1,2}$ and $WEC_3(r)\equiv DEC_3(r)$. Equations \eqref{cond1}-\eqref{cond3} are written in such a way that they are valid for the regions internal and external to the event horizon. Physically, our motivation is to generalize a given GR model where at most the SEC energy condition is the only one to be violated.  In order to establish the energy conditions within the event horizon, because of the change in signature $(-,+,-,-)$, we have to make the following change: $\Theta_{\;\;\nu}^{\mu}=diag\lbrace \rho, -p_r, -p_t, -p_t\rbrace\rightarrow \Theta_{\;\;\nu}^{\mu}=diag\lbrace -p_r, \rho, -p_t(r), -p_t(r)\rbrace$.

In the next section, we will use an algebraic methodology to solve these equations and to obtain new Black-Bounce solutions.

\section{New solutions for black-bounce in $f(R)$ Gravity}\label{sec3}

We will now present three models of new black bounce solutions. The first model is a generalisation of the original Simpson-Visser solution \cite{Visser2}, which has a quadratic action in terms of the curvature scalar $R$. This would be a first approach to checking whether a modified grravity alters the physics given by General Relativity. The second model is a first attempt to obtain a solution that can fulfil the condition $WEC_3$, i.e. $\rho\geq 0$. We will then impose that $\rho=0$ outside the event horizon. In the last model, we'll impose the positivity of the energy density, making $\rho=\rho_0>0$, and thus the condition $WEC_3$ will be satisfied. 
Throughout the article we will use the following metric functions
\begin{eqnarray}
e^{a(r)}=e^{-b(r)}=1-\frac{2m}{\sqrt{r^2+L^2}}\;,\; \Sigma(r)=\sqrt{r^2+L^2}\;.\label{m}
\end{eqnarray}
So the scalar \eqref{k} is
\begin{eqnarray}
K=\frac{4 \left(L^4 \left(-8 M \sqrt{L^2+r^2}+9 M^2+3 r^2\right)+4 L^2 M r^2 \left(2 \sqrt{L^2+r^2}-3 M\right)+3 L^6+12 M^2
   r^4\right)}{\left(L^2+r^2\right)^5}\label{K}\;.
\end{eqnarray}
We see that this scalar is finite throughout space-time, so all the solutions presented here are regular. The event horizon is determined by $e^{a(r_H)}=0$, which results in $r_{H(\pm)}=\pm \sqrt{4m^2-L^2}$. So we have three possibilities: a) $L<2m$, then we have a black bounce with a horizon $r_{H(+)}$ in the positive part of the radial coordinate, a throat at $r=0$, and another horizon $r_{H(-)}$ in the negative part of the radial coordinate; b) $L=2m$ we have a wormhole with a throat at $r=0$ which is an extreme null throat; c) $L>2m$, we have a wormhole with a two-way timelike throat at $r=0$.

\subsection{Quadratic model}\label{subsec1}
We are going to propose the following quadratic model:
\begin{eqnarray}
f(R)=R+a_2R^2\;.\label{mod1}
\end{eqnarray}
In this model we have a quadratic correction, whose coefficient is very small, $a_2<<1$. So this is a correction to GR. 
Through equations \eqref{eq1}-\eqref{eq3}, and the equation of the scalar \eqref{Rscal}, we find the density and pressures
\begin{eqnarray}
&&\rho=-\frac{L^2 \left(2 a_2 L^2 \left(-8 m \sqrt{L^2+r^2}+L^2+15 m^2+r^2\right)+\left(L^2+r^2\right)^2 \left(-4 m
   \sqrt{L^2+r^2}+L^2+r^2\right)\right)}{\kappa ^2 \left(L^2+r^2\right)^5} \,, \\
&&p_r=  -\frac{L^2 \left(6 a_2 L^2 \left(-4 m \sqrt{L^2+r^2}+L^2+3 m^2+r^2\right)+\left(L^2+r^2\right)^3\right)}{\kappa ^2
   \left(L^2+r^2\right)^5} \,, \\
&&p_t=\frac{2 a_2 L^4 \left(-2 m \sqrt{L^2+r^2}+L^2-3 m^2+r^2\right)+L^2 \left(L^2+r^2\right)^2 \left(-m \sqrt{L^2+r^2}+L^2+r^2\right)}{\kappa
   ^2 \left(L^2+r^2\right)^5} \;.
\end{eqnarray}
We can see here that the influence of the quadratic term is given by the coefficient $a_2$, and does not add powers of orders greater than the linear term. If we make $a_2=0$ we fall back on the original Simpson-Visser model of GR. As the asymptotic behaviour at infinity is the same as in GR, that is $\rho\approx p_r\approx -p_t\approx -L^2/(\kappa ^2 r^4)$, which results in $\rho+p_r<0$ in this region, violating $NEC$, implying the violation of all the energy conditions, just as in the case of GR. Adding a quadratic term did not lead to any physical change. It may be that there is some non-linear contribution to $f(R)$ that implies the non-violation of $NEC$, but in general we don't know how to arrive at this condition directly.

\subsection{Zero energy density model}\label{subsec2}

\begin{figure}[H]

\centering
\includegraphics[width=0.4\textwidth]{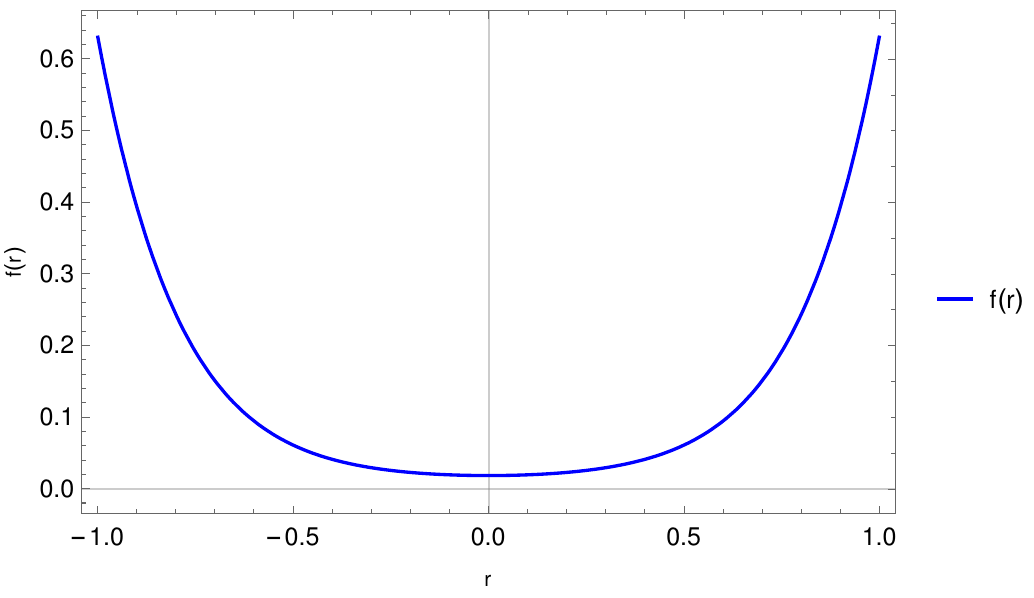}
 \qquad
\includegraphics[width=0.4\textwidth]{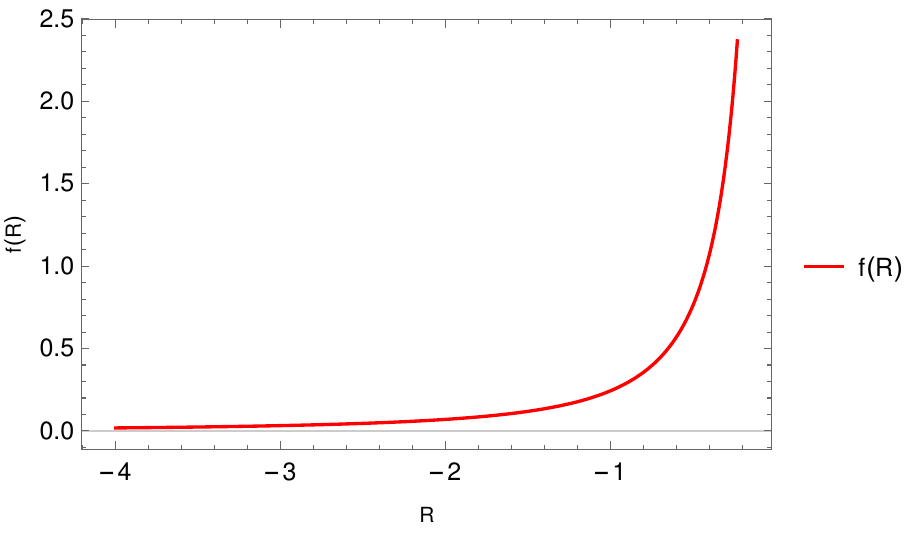}
\caption{\scriptsize{Graphic representation of the functions $f(r)\times r$ (blue) and $f(R)\times R$ (red), to $m=1$ and $L=1$.} }
\label{f(R)1}
\end{figure}

\begin{figure}[H]

\centering
\includegraphics[width=0.4\textwidth]{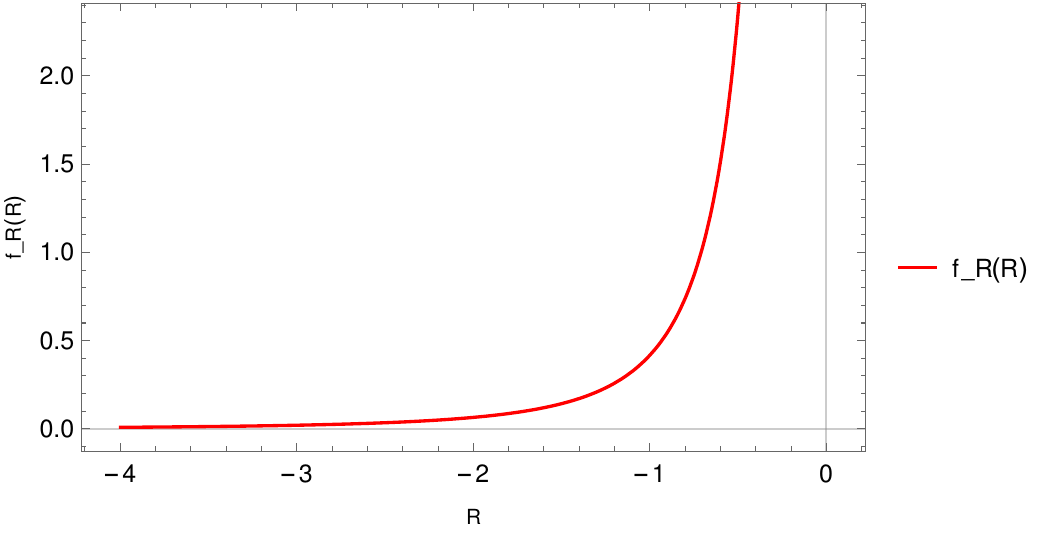}
 \qquad
\includegraphics[width=0.46\textwidth]{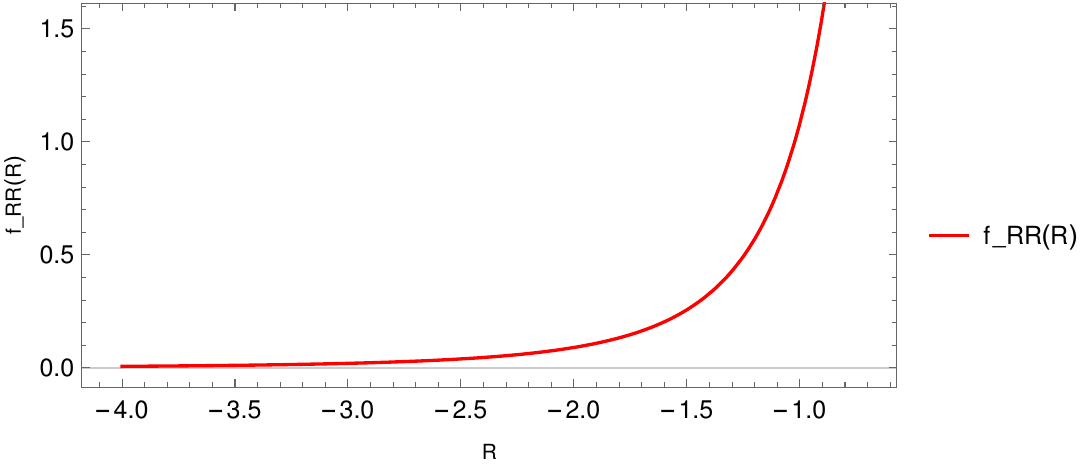}
\caption{\scriptsize{Graphic representation of the functions $f_{R}(R)\times R$ (blue) and $f_{RR}(R)\times R$ (red), to $m=1$ and $L=1$.} }
\label{fRM1}
\end{figure}

We are now interested in obtaining a nonlinear $f(R)$ function for this solution satisfying $WEC_3\equiv DEC_3$. We do this by setting the left side of \eqref{eq1} equal to zero. According to eq.\eqref{conddec3}, using the definition of $b(r)$ and integrating to get $f(r)$,
we obtain,
\begin{eqnarray}
f(r)=e^{-4\frac{\sqrt{L^2+r^2}}{m}}\left(L^2+r^2\right)^{15/2}\,. \label{fr1}
\end{eqnarray}
Note that here \eqref{fr1} is a function of the coordinate $r$, not the scalar of curvature $R$. However, we can plot the parameterized graphic of $f(r)\times r$ and $f(R )\times R$ numerically calculated using the curvature scalar \eqref{Rscal} and $f_R=\frac{df}{dr}\left(\frac{dR}{dr}\right)^{-1}$, as can be seen in Fig.\,\ref{f(R)1},
showing clearly that the theory is non-linear in  $f(R)$.  We also see that this model does not violate the condition of non-existence of phantoms or tachyons, since its first and second derivatives are positive, as can be seen in Fig.\,\eqref{fRM1}. This indicates that the obtained model presents stability.

\par
\begin{figure}[H]
\centering
\qquad
\includegraphics[width=0.45\textwidth]{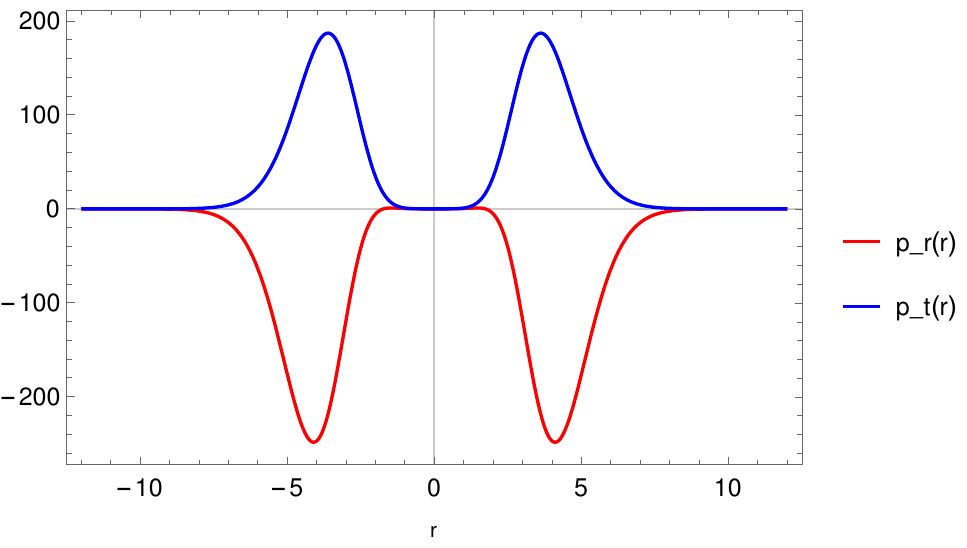}

\caption{\scriptsize{Graphic representation for radial pressure, \eqref{prx}, and tangential pressure, \eqref{ptx}.} }
\label{fluidprimeira}
\end{figure}

\begin{figure}[H]
\centering

\subfloat[Inside the horizon.]{ \label{subfig:nec1d}
\includegraphics[width=0.4\textwidth]{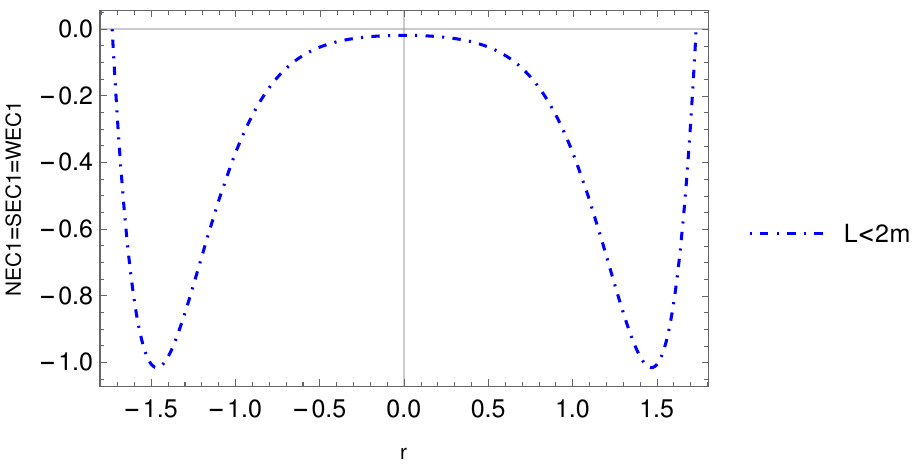}
}
 \qquad
\subfloat[Outside the horizon and $L>2m$ and $L=2m$.]{ \label{subfig:nec1f}
\includegraphics[width=0.4\textwidth]{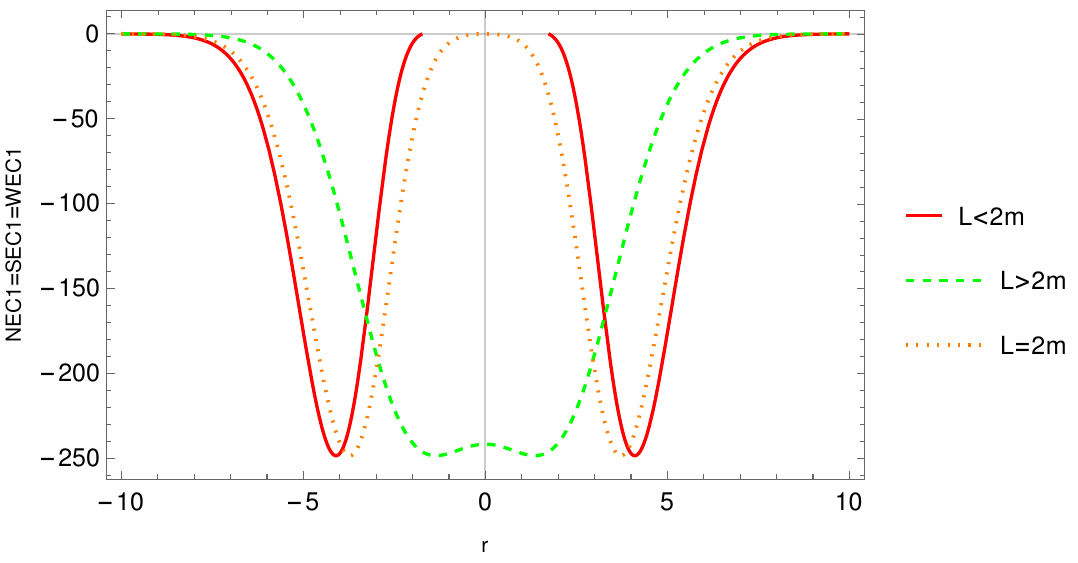}
}\\
\includegraphics[width=0.4\textwidth]{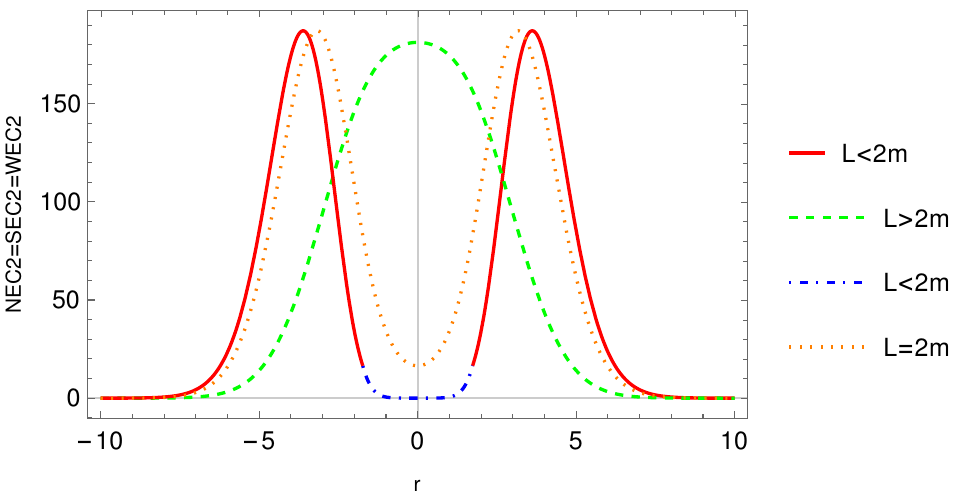}
\qquad
\includegraphics[width=0.4\textwidth]{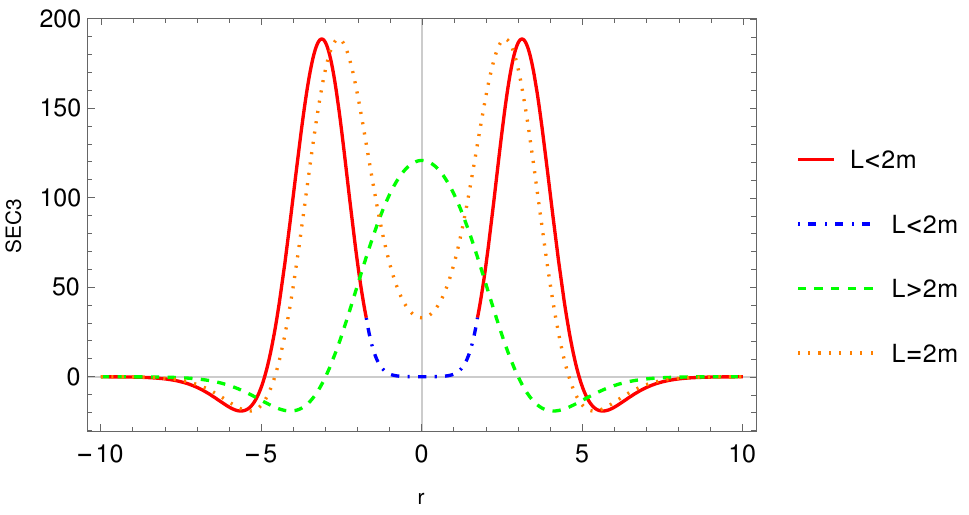}\\
\subfloat[Inside the horizon.]{ \label{subfig:dec1d}
\includegraphics[width=0.4\textwidth]{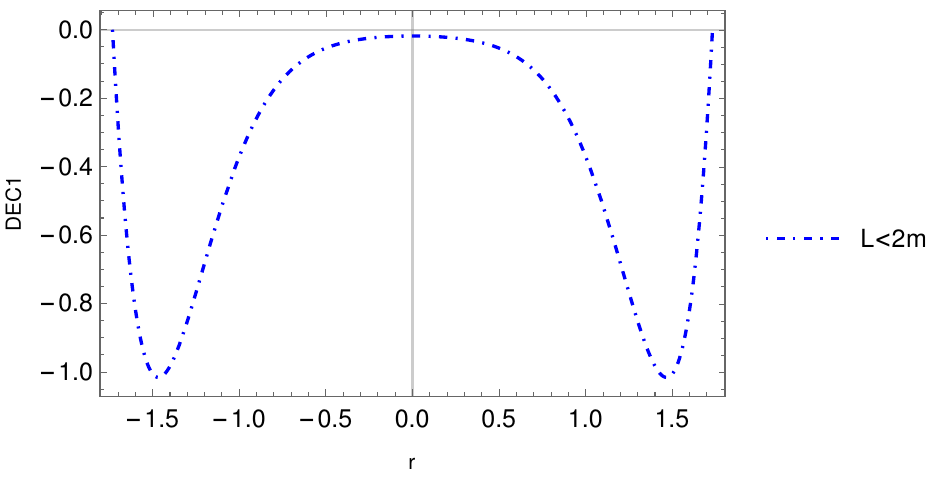}
}
\qquad
\subfloat[Outside the horizon and $L>2m$ and $L=2m$.]{ \label{subfig:Dec1f}
\includegraphics[width=0.4\textwidth]{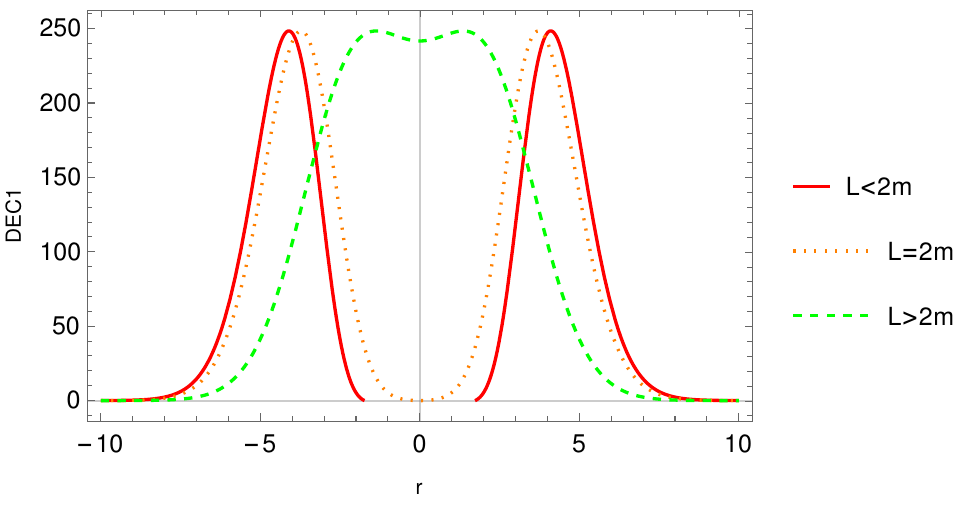}
}\\

\includegraphics[width=0.4\textwidth]{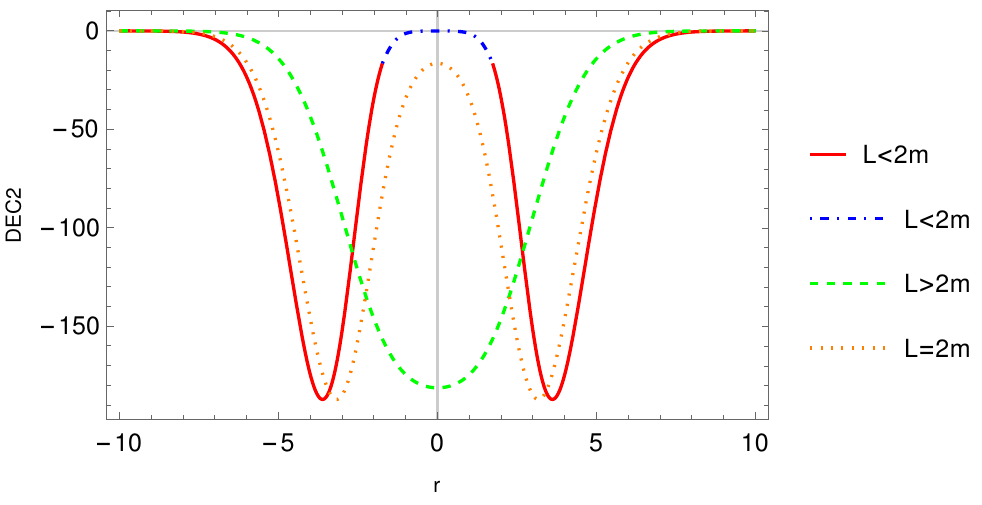}
\qquad
\includegraphics[width=0.4\textwidth]{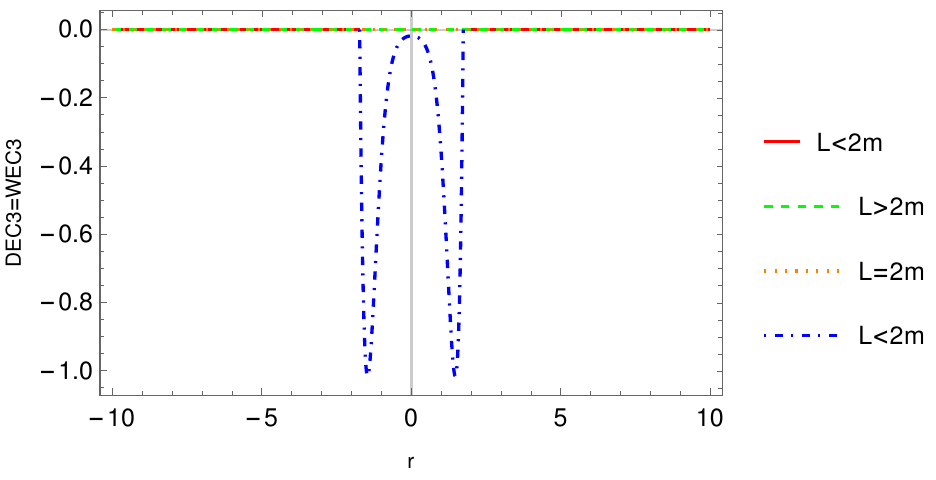}
\caption{\scriptsize{Graphic representation of the energy conditions for spacetime obtained for the model given by the equation \eqref{fr1}, in the region where $t$ is timelike(red) and region where $t$ is spaceliket(blue with dash-dot) for $m=1$ $L=1$ and for spacetime with $L=2$(orange with dot-dot) and $L=4$(green with dash-dash).} }
\label{ECquad2}
\end{figure}

 
The expressions for $p_r$ e $p_t$ have been calculated for the mode \eqref{fr1} using the equations \eqref{eq2} and \eqref{eq3},
\begin{eqnarray}
&& p_r=-\frac{e^{-4\frac{\sqrt{L^2+r^2}}{m}}}{m}\left[\left(L^2+r^2\right)^7\left(L^2+r^2-2m\sqrt{L^2+r^2}\right)\right]\,,\label{prx}\\
&&p_t=\frac{3}{2}e^{-4\frac{\sqrt{L^2+r^2}}{m}}\left(L^2+r^2\right)^{15/2}\,,\label{ptx}
\end{eqnarray}
and plotted in Fig.\,\ref{fluidprimeira} where we can see that there is no divergence associated with the fluid, both in regions external and internal to the event horizon (if any).  
Note that here, for any value of $r$, $p_r<0$ and $p_t>0$, which indicates, as we will see, that $NEC_1$ is violated throughout \textcolor{red}{all} spacetime. For regions inside the event horizon, in which case $L<2m$, the coordinate $t$ becomes spacelike and therefore $\Theta^{\mu}_{\,\,\nu}=diag(-p_r , \rho, -p_t, -p_t)$. Thus, $\rho$ inside the event horizon is non-zero, however, $p_r=0$ for regions inside the horizon.
The energy conditions are obtained from \eqref{cond1}-\eqref{cond3} using \eqref{prx} and \eqref{ptx} and are represented in Fig.\ref{ECquad2} for regions inside and outside the horizon of events, in which case $L<2m$, and for cases where $L>2m$ and $L=2m$. For $NEC_1$ and $DEC_1$, the region inside the event horizon is shown separately due to its graphic proportionality in relation to the regions outside the horizon, Fig.\,\eqref{subfig:nec1d} and \eqref{subfig:dec1d}

Here, then, we have that $NEC_1$ is violated in all spacetime for all cases, while $NEC_2$ is satisfied in all spacetime for all cases. $SEC_3$ is satisfied in a small region $r_{H_+}<r<r_1$ and $-r_1<r<r_{H_-}$ for the case of a regular black hole, in which case the event horizon $r_{H_{\pm}}$ is present($L<2m$), and violated within this horizon $r_{H_-}<r<r_{H_+}$, as in GR for a regular black hole. For $L>2m$ and $L=2m$, $SEC_3$ is satisfied for regions close to the origin of the coordinate $r$, that is, $-r_1<r<r_1$ and violated for $r>r_1$ and $- r_1>r$. $DEC_1$ is violated within the event horizon $r_{H_-}<r<r_{H_+}$ and satisfied outside the horizon for a regular black hole; for both cases $L>2m$ and $L=2m$ the $DEC_1$ is satisfied. $DEC_2$ is violated in all three cases. $DEC_3$, and therefore $WEC_3$, by definition is satisfied outside the event horizon and violated inside the event horizon for $L<2m$. For the other two cases it is satisfied since $\rho= 0$.

\subsection{Positive density model}\label{subsec3}
Now we want to show that there is a huge difference between the models that can be obtained from $f(R)$ gravity and those derived from GR. Here we can formulate a model that satisfies $NEC_1$ and also satisfies $WEC_3$ together. To do this, we need to manipulate the equations of the theory appropriately. 
First, let us subtract \eqref{eq1} to \eqref{eq2}, to find $p_r$:
\begin{eqnarray}
p_r=\frac{2 m f_R''(r)}{\kappa ^2 \sqrt{L^2+r^2}}-\frac{f_R''(r)}{\kappa ^2}+\frac{4 L^2 m f_R(r)}{\kappa ^2
   \left(L^2+r^2\right)^{5/2}}-\frac{2 L^2 f_R(r)}{\kappa ^2 \left(L^2+r^2\right)^2}-\rho\,.\label{pr1}
\end{eqnarray}
Now, to ensure that the $WEC_3$ condition is satisfied, we impose $\rho=\rho_0>0$. Then we solve \eqref{eq3} to determine $p_t$
\begin{eqnarray}
p_t=\frac{f(r) \left(L^2+r^2\right)^{5/2}+2 \left(L^2+r^2\right) \left(\left(L^2+r^2\right) f_R''(r) \left(\sqrt{L^2+r^2}-2 m\right)+r
   \sqrt{L^2+r^2} f_R'(r)\right)+4 L^2 m f_R(r)}{2 \kappa ^2 \left(L^2+r^2\right)^{5/2}}\,.\label{pt1}
\end{eqnarray}
Continuing, we use the identity $f_R=(df/dr)/(dR/dr)$ and we insert the previous expressions in \eqref{eq1}, integrating the result to determine $f$:
\begin{eqnarray}
f(r)=-e^{-\frac{4 \sqrt{L^2+r^2}}{m}} \left[f_1 \left(L^2+r^2\right)^{15/2}+2 \kappa ^2 \rho_0 e^{\frac{4 \sqrt{L^2+r^2}}{m}}\right]\,,\label{f1}
\end{eqnarray}
where $f_1>0$. Now we have all the functions we are looking for. We can show that 
\begin{eqnarray}
\rho+p_r=\frac{f_1 \left(L^2+r^2\right)^{15/2} e^{-\frac{4 \sqrt{L^2+r^2}}{m}} \left(\sqrt{L^2+r^2}-2 m\right)^2 \left(4 \sqrt{L^2+r^2}-15
   m\right)}{\kappa ^2 m \left(-23 m \sqrt{L^2+r^2}+4 L^2+30 m^2+4 r^2\right)}\,.\label{condp}
\end{eqnarray}
We can now see that the asymptotic behaviour at infinity is $\rho + p_r\approx f_1 r^{16} e^{-\frac{4 r}{m}}/(m\kappa^2)>0$. So $NEC_1$ is satisfied, and since we chose $\rho=\rho_0>0$, $WEC_3$ is also satisfied, In GR, we can't have this result where two of the partial energy conditions are satisfied at the same time. This was only possible because the energy density, radial pressure and tangential pressure were added with new terms from the GR modification.

\section{Conclusion}\label{sec4}

 In this article we have exploited the possibility to obtain black-bounce solutions using $f(R)$ theories. Black-bounce solutions include regular black holes as well as wormhole configurations which can be traversable in one or in both directions. These solutions combines the structure of well-known regular black holes, like the Bardeen and Hayward ones, with those of wormholes like the Ellis-Bronnikov wormhole. In General Relativity, the implementation of the black-bounce solutions is possible but at the cost of violation of energy conditions. Possibly, this may be problematic from the point of view of the stability of the configuration. The non-linear generalization of GR given by the class of $f(R)$ theories may open new possibilities, and the goal of our analysis was to investigate how the usual GR scenario can change using those theories.

We began by analysing a quadratic model of the function $f(R)$. This model, $f(R)=R+a_2R^2$, is a correction of GR, taking $a_2<<1$. This could give us an idea of how the addition of non-linear terms affects the physics of black bounce solutions. In this model, we can always take the limit $a_2\rightarrow 0$, falling back on GR. We can also examine how the coefficient $a_2$ affects the material content, thus modifying the energy conditions of the system. In our analysis, it can be seen that the inclusion of the quadratic term did not asymptotically affect the energy conditions at infinity of the radial coordinate, thus maintaining the same result as GR. This is an unusual result, as we expected a quadratic term to add higher powers of the radial coordinate to the expressions for the energy density and the radial and tangential pressures. This is not the case.

In the second model, we first tried to ensure that the condition $WEC_3=\rho\geq 0$ was fulfilled. To this end, we imposed $\rho=0$, and integrated the equation of motion. The result is that we can obtain a solution with zero energy density, and which has two horizons, one in the positive part, and the other in the negative part of the radial coordinate.This did not occur in the case of GR, where the same imposition led to a solution without horizons, see \cite{manuelmarcos}. In our case, the energy conditions are satisfied in some intervals of the radial coordinate, which was not the case in GR. We also showed the property that the function $f(R)$, its first and second derivatives, are positive. This may indicate stability with respect to small perturbations.

Finally, we propose a model in which the energy density is a positive constant, which leads us to the condition $\rho+p_r>0$, thus satisfying two partial energy conditions, $WEC_3$ and $NEC_1$. You can not obtain $\rho+p_r>0$ in GR without changing the expression of $\Sigma$. This was only possible here because of the non-linear contributions added to the radial and tangential pressures from the $f(R)$ function. 

The $f(R)$ theories have some conditions to be considered viable. In particle, in order to have positive sound speed and, at the same time, not containing tachyons or ghost, the first and second derivatives of the function with respect to the curvature scalar must be positive. At least in some cases analysed here, the specific $f(R)$ model satisfies these requirements. This an encouraging aspect of our results. On the other hand, the stability of the final configuration remains an open question being a highly non-trivial problem, in part because of the existence of a throat in the wormhole and black hole solutions \cite{st1,st2,st3}. 

 The next steps related to this work are: to verify the type of matter that models our solutions; to analise the possible black-bounce shadows; to obtain the thermodynamics of solutions; and to analise the absorption and scattering of the scalar field for this type of solution.

\vspace{1cm}

{\bf Acknowledgement}: MER thanks CNPq for partial financial support. JCF thanks CNPq and FAPES for financial support.



\end{document}